\documentclass[twocolumn]{aastex62}

\submitjournal{ApJ}
\accepted{November 18, 2019}

\usepackage{epstopdf}
\usepackage{graphicx}
\usepackage{threeparttable}
\usepackage{longtable}

\begin{document}

\title{Star-forming Clumps in Local Luminous Infrared Galaxies}

\author{K. L. Larson}
\affiliation{Department  of  Astronomy,  California  Institute  of  Technology,  1200  E. California  Blvd.,  Pasadena,  CA  91125,  USA}
\affiliation{IPAC, California Institute of Technology, 1200 E. California Blvd., Pasadena, CA 91125, USA}

\author{T. D\'iaz-Santos}%Tanio D\'iaz-Santos
\affiliation{N\'ucleo  de  Astronom\'ia  de  la  Facultad  de  Ingenier\'ia,  Universidad  Diego  Portales,  Av.  Ej\'ercito  Libertador  441,  Santiago,  Chile}

\author{L. Armus}
\affiliation{IPAC, California Institute of Technology, 1200 E. California Blvd., Pasadena, CA 91125, USA}

\author{G. C. Privon} %George Privon
\affiliation{Department  of  Astronomy,  University  of  Florida,  211  Bryant Space  Sciences  Center,  Gainesville,  FL  32611,  USA}

\author{S. T. Linden}
\affiliation{Department  of  Astronomy,  University  of  Virginia,  Charlottesville,  VA  22904,  USA}
\author{A. S. Evans}
\affiliation{Department  of  Astronomy,  University  of  Virginia,  Charlottesville,  VA  22904,  USA}
\affiliation{National Radio Astronomy Observatory, 520 Edgemont Road, Charlottesville, VA 22903 USA}

\author{J. Howell}
\affiliation{IPAC, California Institute of Technology, 1200 E. California Blvd., Pasadena, CA 91125, USA}

\author{V. Charmandaris}
\affiliation{Department  of  Physics,  University  of  Crete,  GR-71003,  Heraklion,  Greece}
\affiliation{Institute of Astrophysics, Foundation for Research and Technology - Hellas, Heraklion, GR-70013, Greece}
\author{V. U}
\affiliation{Department of Physics and Astronomy, 4129 Frederick Reines Hall, University of California, Irvine, CA 92697, USA}

\author{D. B. Sanders}
\affiliation{Institute for Astronomy, University of Hawaii, 2680 Woodlawn Drive, Honolulu, HI 96822}

\author{S. Stierwalt}
\affiliation{IPAC, California Institute of Technology, 1200 E. California Blvd., Pasadena, CA 91125, USA}
\affiliation{Department of Physics, Occidental College, 1600 Campus Road, Los Angeles, CA 90041}
\author{L. Barcos-Mu\~noz}
\affiliation{Joint ALMA Observatory, Alonso de C\'ordova 3107, Vitacura, Santiago, Chile}
\affiliation{National Radio Astronomy Observatory, 520 Edgemont Road, Charlottesville, VA 22903 USA}

\author{J. Rich}
\affiliation{Carnegie Observatories, 813 Santa Barbara St., Pasadena, CA 91101}

\author{A. Medling} 
\affiliation{Ritter Astrophysical Research Center, University of Toledo, Toledo, OH 43606, USA}
\affiliation{Department  of  Astronomy,  California  Institute  of  Technology,  1200  E. California  Blvd.,  Pasadena,  CA  91125,  USA}

\author{D. Cook} 
\affiliation{IPAC, California Institute of Technology, 1200 E. California Blvd., Pasadena, CA 91125, USA}
%\affiliation{Department  of  Astronomy,  California  Institute  of  Technology,  1200  E. California  Blvd.,  Pasadena,  CA  91125,  USA}

\author{A. Oklop\^ci\'c}%Antonija Oklop\^ci\'c
\affiliation{Department  of  Astronomy,  California  Institute  of  Technology,  1200  E. California  Blvd.,  Pasadena,  CA  91125,  USA}
\affiliation{Institute for Theory and Computation, Harvard-Smithsonian Center for Astrophysics
60 Garden Street, MS-51, Cambridge, Massachusetts 02138, USA}

\author{E. J. Murphy} 
\affiliation{National Radio Astronomy Observatory, 520 Edgemont Road, Charlottesville, VA 22903 USA}

\author{P. Bonfini}
\affiliation{Institute of Astrophysics, Foundation for Research and Technology - Hellas, Heraklion, GR-70013, Greece}
\affiliation{Institute for Astronomy, Astrophysics, Space Applications \& Remote Sensing, National Observatory of Athens, P. Penteli, GR-15236 Athens, Greece}

%\author{others}

\begin{abstract}
We present $HST$ narrow-band near-infrared imaging of Pa$\alpha$ and Pa$\beta$ emission of 48 local Luminous Infrared Galaxies (LIRGs) from the Great Observatories All-Sky LIRG Survey (GOALS). These data allow us to measure the properties of 810 spatially resolved star-forming regions (59 nuclei and 751 extra-nuclear clumps), and directly compare their properties to those found in both local and high-redshift star-forming galaxies. 
We find that in LIRGs, the star-forming clumps have radii ranging from $\sim90-900$~pc and star formation rates (SFRs) of $\sim1\times10^{-3}$ to 10~M$_\odot$yr$^{-1}$, with median values for extra-nuclear clumps of 170~pc and 0.03~M$_\odot$yr$^{-1}$.
The detected star-forming clumps are young, with a median stellar age of $8.7$~Myrs, and a median stellar mass of $5\times10^{5}$~M$_\odot$. 
The SFRs span the range of those found in normal local star-forming galaxies to those found in high-redshift star-forming galaxies at $\rm{z}=1-3$. 
The luminosity function of the LIRG clumps has a flatter slope than found in lower-luminosity, star-forming galaxies, indicating a relative excess of luminous star-forming clumps. In order to predict the possible range of star-forming histories and gas fractions, we compare the star-forming clumps to those measured in the MassiveFIRE high-resolution cosmological simulation. The star-forming clumps in MassiveFIRE cover the same range of SFRs and sizes found in the local LIRGs and have total gas fractions that extend from 10 to 90\%. If local LIRGs are similar to these simulated galaxies, we expect future observations with ALMA will find a large range of gas fractions, and corresponding star formation efficiencies, among the star-forming clumps in LIRGs.

\end{abstract}

%\nopagebreak

\section{Introduction}

Star formation processes are central to understanding the buildup of mass in galaxies over time. 
Star formation occurs over a large range in physical scale, from pc-sized molecular clouds to large kpc-sized star-forming regions in the disks and nuclei of galaxies (e.g. Kennicutt \& Evans 2012).
Connecting star formation over these vastly different scales is difficult, but essential  
to understanding fundamental observable signatures of galaxy evolution, such as the main sequence of star-forming galaxies, and the role of mergers (Daddi et al. 2007, Elbaz et al. 2007, 2011, Rodighiero et al. 2011). Additionally, resolving star-forming regions and linking emission from young stars to the physical conditions and gas supply is critical to understanding the ubiquity and genesis of the Kennicutt-Schmidt law (KS, Schmidt 1959, Kennicutt 1998) and the variation of star formation efficiency across galaxies and galaxy types.

Recently, a number of surveys have found that high-z star-forming galaxies tend to display turbulent, clumpy disks with extreme star-forming clumps, having  masses of $\sim 10^{8-9}$~M$_\odot$ and sizes of 0.5 -- 5~kpc (Elmegreen et al. 2004, 2009, Daddi 2010, Genzel et al. 2008, Livermore et al. 2015).  These clumps are $\sim100\times$ larger than typical giant molecular cloud (GMCs) locally. Most galaxies studied in detail at high-z host a few large, kpc-sized star-forming clumps. In contrast, local star-forming galaxies contain hundreds of smaller HII regions (e.g. Cowie, Hu, \& Songaila 1995, Wisnioski et al. 2012, Cook et al. 2016), suggesting a fundamental shift in the way stars form in galaxies at these early epochs. 

One possible interpretation is that these large star-forming clumps are a result of higher gas fractions and increased star formation efficiency (Tacconi et al. 2008, Jones et al. 2010). Simulations by Tamburello et al. (2015) support this hypothesis, showing that simulated galaxies with high gas fractions are more likely to have large star-forming clumps than galaxies with lower gas fractions. Furthermore, these massive clumps appear to be present in both interacting and non-interacting high-z galaxies. 
%-------------

Turbulence and high gas fractions have both been invoked to explain the extreme properties of star-forming clumps in high-z galaxies (Dekel, Sari, \& Ceverino 2009,  Forster Schreiber et al. 2011, Soto et al 2017).  However, published studies often have limited spatial resolution and sensitivity, dropping normal star-forming regions well below the detection limits, and making it difficult to resolve sources on the scales of traditional HII regions. The true luminosity function and size distributions of star-forming regions in high-z galaxies is therefore still highly uncertain.
 
%Lensing studies look at some z$\sim$1 sources and with higher resolution.
Recent progress has been made using strong gravitational lensing of z = 1--3 galaxies, which has allowed for clump studies at resolutions down to 100pc. Livermore et al. (2012, 2015) found that the average star formation rate density (SFR/kpc${^{-2}}$) of star-forming clumps in lensed, star-forming galaxies increases with redshift. Other authors have studied the effects of resolution by smoothing intrinsically high-res data to lower resolution to mimic the effects of observing high-z galaxies without the benefit of lensing.  Fisher (2017), in a study of $z\sim0.1$ galaxies, found that the size of the detected clumps grew and the number of clumps fell as a larger fraction of smaller clumps got incorporated in the beam. Cava et al. (2018) directly tested the effects of resolution on clump properties by studying multiple lensed images of the same $z=0.4385$ galaxy at different magnifications. The different magnifications allow them to study the same galaxy at resolutions from 30 to 300~pc and show that lower resolution, not surprisingly, causes the clumps size and masses to be systematically overestimated. 

These studies illustrate the need to study star-forming clumps in actively star-forming galaxies at high spatial resolution and with enough sensitivity to probe the faint end of the luminosity function. It is also important to be able to separate clumps from multiple nuclei, and disk clumps from those in and around the nuclei, especially in galaxies undergoing interactions and mergers, which provide the trigger for the enhanced star formation, since in these systems circumnuclear clumps can be quite complex and the differences between nuclei and luminous star-forming clumps can become blurred.

%------------------------------------------------------------------------

Luminous Infrared Galaxies (LIRGs), with thermal IR[8-1000~$\mu$m] dust emission in excess of 10$^{11}$~L$_\odot$, are an ideal laboratory for studying resolved star formation in the local Universe. 
The bolometric luminosity of LIRGs is dominated by massive bursts of star formation, showing also a wide range of contributions from active galactic nuclei (AGN - see Petric et al. 2011; Stierwalt et al. 2013, 2014). Multi-wavelength observations have shown that local LIRGs are a mixture of single disk galaxies, interacting systems, and advanced mergers, exhibiting enhanced SFRs and AGN activity compared to less luminous and non-interacting galaxies (c.f. Sanders \& Mirabel 1996, Stierwalt et al., 2015, Larson et al., 2016).  By comparing star formation in LIRGs to normal (non-interacting and L$_{\rm{IR}}<10^{11}$) low redshift galaxies, high redshift galaxies, and sophisticated hydrodynamical simulations, we can better understand how global galaxy properties and environment influence star formation on smaller scales.

Because LIRGs are, by their very nature, dusty, it is important to study star formation in these systems in the infrared, where obscuration is  minimized while still achieving high spatial resolution. 
%Why Pbeta and Palpha
In star-forming galaxies the hydrogen recombination emission lines are used extensively as SFR estimators, since their fluxes provide a straightforward measure of the number of ionizing photons produced by massive O/B stars. However, the central kpc of LIRGs are often heavily enshrouded and can be subject to extremely large visual extinctions, A$_V$ $>$ 10 (Garc\'ia-Mar\'in et al. 2009, Stierwalt et al. 2013, Piqueras L\'opez, J. et al. 2013), making the optical H$\alpha$ and H$\beta$ lines poor tracers of the star formation (Armus et al. 1989).  
The extinction throughout the galaxy is much less severe in the near-IR (NIR; 1.6~$\mu$m $\sim$ 0.2 × A$_V$) and lines such as Pa$\alpha$, Pa$\beta$ or Br$\gamma$ can provide a much more complete picture of the obscured star formation.

Integral field spectroscopy (IFS) observations on large, ground based telescopes can be used to simultaneously measure the distribution and luminosity of star-forming clumps, as well as an estimate of the extinction from ratios of the Paschen and Bracket recombination lines (e.g., U et al. 2018), but currently these studies have extremely limited fields of view.  For example, in a detailed study of 17 (U)LIRGs using SINFONI on the VLT, Piqueras L\'opez et al (2016) found Pa$\alpha$ luminosities and SFRs for some clumps that rival those seen at high-redshift in the inner 8''x8'' (typically 3~kpc) of their galaxies. Deep, high-resolution, wide field of view studies are better suited to measuring the star formation across entire merging systems as a function of merger stage and disk properties, and for comparison to galaxies at high redshift.
 
Here, we present Hubble Space Telescope ($HST$) near-infrared, narrow-band imaging study of the properties of 810 star-forming clumps in a sample of 48 local LIRGs.  The sample galaxies cover a range of merger stages, from undisturbed non-interacting galaxies, to highly disturbed late-stage mergers. The spatial resolution of $HST$ enables us to quantify the fraction of star formation occurring in clumps versus faint, diffuse emission, and to measure the size, luminosities, and spatial distribution of star-forming clumps on sub-kpc scales. 

In this paper we present the global properties of star-forming clumps in local LIRGs, measuring the SFRs, sizes, ages, and masses of star-forming clumps via the high-resolution Pa$\alpha$ and Pa$\beta$ $HST$ images. In a follow-up paper (Paper-II), we will investigate how the clump properties change with individual galaxy properties like the global SFR, sSFR, merger stage, and as a function of radial distance in each galaxy.  In section 4.1 we measure the luminosity function (LF) of the star-forming clumps and compare the slope of the clump LF in GOALS to that found in local galaxies by Cook et al. (2016). In section 4.2 we investigate the size and SFR of the clumps and compare them to both high redshift ($1<z<4$) star-forming clumps from Livermore et al. (2012, 2015) and to star-forming clumps in z=0 normal (non-interacting and L$_{\rm{IR}}<10^{11}$) galaxies in the SINGS sample. Finally, in section 4.3 we compare our the GOALS clumps to star-forming regions in galaxies in MassiveFIRE, a high-resolution cosmological hydrodynamical simulation.

\section{Sample Selection and Observations}

The Great Observatories All-sky LIRG Survey (GOALS:  Armus et al. 2009) is a complete galaxy sample that comprises the 201 LIRG systems (z $< 0.088$) included in the IRAS Revised Bright Galaxy Sample (RBGS, Sanders et al. 2003), and is aimed at measuring the physical properties of local LIRGs across the electromagnetic spectrum using a broad suite of ground and space-based observatories. 

Here, we present and analyze Pa$\alpha$ and Pa$\beta$ $HST$ images of a sample of 48 LIRGs in GOALS with L$_{\rm{IR}}$s that range from 10$^{11.2}$ to 10$^{12.3}$. The Pa$\alpha$ data (27 systems) are $HST$ archival images from PID 10169. These galaxies were selected to have a redshift range of 0.0093 $\le$ z $\le$ 0.0174
so that the Pa$\alpha$ emission line lies within the $HST$ NICMOS F190N narrow-band filter (Alonso-Herrero 2006).
Galaxies in the Pa$\beta$ sample were selected to have redshifts between 0.0225 $\le$ z $\le$ 0.0352 so that the Pa$\beta$ emission line lies within the $HST$ WFC3 F132N narrow-band filter. Since most of the galaxies in the Pa$\alpha$ sample were isolated galaxies and early stage mergers, mid to late-stage mergers were chosen for the expanded Pa$\beta$ sample. The Pa$\beta$ sample is therefore not a flux-limited sample, as it does not contain all the galaxies in the redshift range that could be observed. The combination of the Pa$\alpha$ and Pa$\beta$ samples contain 12 non-interacting galaxies, 17 early-stage mergers, and 17 mid and late-stage mergers providing a full representation of the entire interaction sequence.  When all the galaxies in pairs are counted, there are a total of 59 individual galaxies in the sample.

\subsection{Observations}

\begin{figure*}
	\centering
	\includegraphics[width=0.98\textwidth]{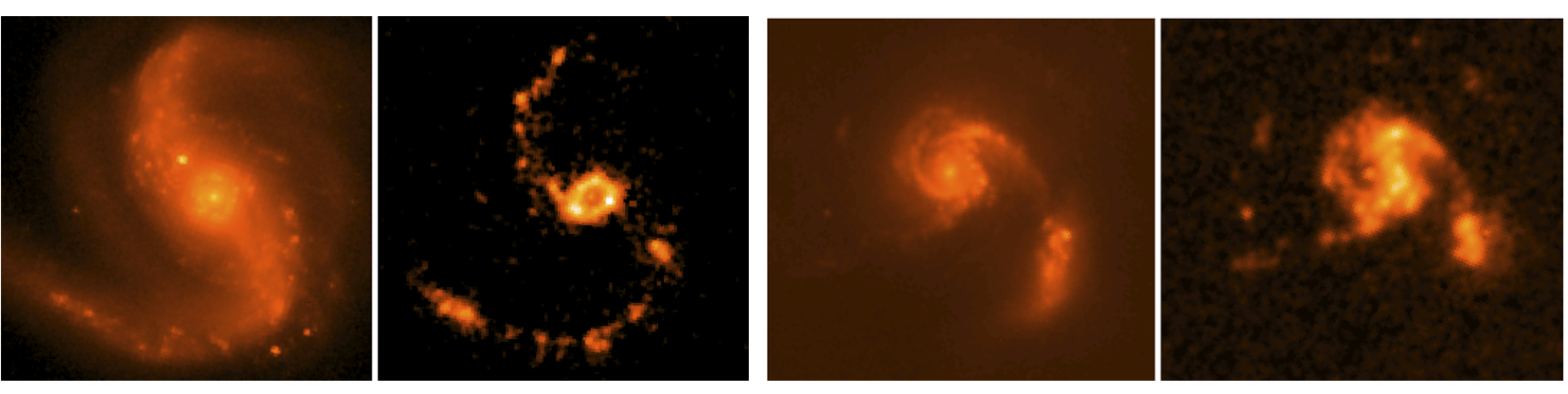}
	\caption{Left Figure: HST/WFC3 F110W (left panel) and continuum subtracted Pa$\beta$ image (right panel) of NGC~6786. Right Figure: HST/WFC3 F110W (left panel) and continuum subtracted Pa$\beta$ image (right panel) of NGC~6090. Clumps of star formation and diffuse emission are visible in both NGC~6786 and the late-stage merger NGC~6090. In particular, NGC~6786 has a ring of star formation around the nucleus, while NGC~6090 contains a bright extra-nuclear starburst between the two interacting nuclei. }
	\label{FIG:PaImg}
\end{figure*}

%cycle 23 
The Pa$\beta$ observations were taken with the $HST$ WFC3 camera (Project ID: 13690; PI: D\'iaz-Santos), which has a plate scale of 0\farcs12~pixel$^{-1}$ using the F130N and F132N narrow-band filters and the F110W broad-band filter. 
The total integration times were 120s for each broadband filter and about 1100s (19~min) for each narrowband filter. 
The F130N filter is used to measure the underlying continuum emission next to the line. The 2' field of view (FoV) of WCF3-IR allows for galaxies in a pair to be observed simultaneously. Each observation was again divided into three individual exposures with offsets of 5 pixels.

%Archival
%https://archive.stsci.edu/proposal_search.php?id=10169&mission=hst
The archival Pa$\alpha$ observations were taken with the $HST$ NIC2 camera, which has a plate scale of 0\farcs076~pixel$^{-1}$ and a 19\farcs2 FoV using the F187N and F190N narrow-band filters. This FoV allows for detection of P$\alpha$ emission up to a radial distance $\sim$ 5~kpc from the galaxy nucleus. From our Pa$\beta$ data, which covers a much larger FoV, we find that most of the Paschen emission is indeed contained within 5~kpc from the galaxy nucleus.
For objects with redshifts between $0.0093$ and $0.0174$ the Pa$\alpha$ emission line, at 1.87~$\mu$m, is captured with the F190N filter, while the adjacent filter, F187N, is used for continuum subtraction.
The typical total integration times were
900-950~s for each narrow-band filter. Each observation was divided into three individual exposures with offsets of 5 pixels. 
For details on the data reduction of the Pa$\alpha$ observations we refer the reader to Alonso-Herrero et al. 2006. Observational details are given for every galaxy in our combined Pa$\alpha$ and Pa$\beta$ sample in Table \ref{TAB:Observations}.

%%-------------------------------
%Galaxy properties Table
\begin{deluxetable*}{lllll}
\tablecolumns{5} 
\tablecaption{Observations}
 \tablehead{
 \colhead{Galaxy Name} & \colhead{IRAS Name} & \colhead{z} & \colhead{Filter} & \colhead{PIDs}
 }
\startdata
NGC 0023   & IRAS F00073+2538  &0.015231 &F187N, F190N & 10169   \\
MCG -02-01-051 & IRAS F00163-1039  & 0.027152 & F130N, F132N &  13690    \\
MCG +12-02-001  & IRASF 00506+7248 &0.015698 & F187N, F190N & 10169  \\
NGC 0633   & IRASF 01341-3735 & 0.017314 & F187N, F190N & 10169   \\
IC 0214 & IRAS F02114+0456  & 0.030224 & F130N, F132N &  13690    \\
UGC 1845   & IRASF 02208+4744 &0.015607 & F187N, F190N &  10169    \\
 ..... & IRAS F02437+2122  & 0.023306 & F130N, F132N &  13690    \\
 ..... & IRAS F03217+4022  & 0.023373 & F130N, F132N &  13690    \\
......   & IRAS 03582+6012  & 0.030011 & F130N, F132N &  13690    \\
NGC 1614   & IRASF 04315-0840 & 0.015938 & F187N, F190N &  10169   \\
......   & IRAS 05129+5128  & 0.027432 & F130N, F132N &  13690    \\
UGC 3351   & IRASF 05414+5840 & 0.01486 & F187N, F190N &  10169   \\
NGC 2369   & IRASF 07160-6215 & 0.010807& F187N, F190N &  10169   \\
NGC 2388   &  IRASF 07256+3355 & 0.01379& F187N, F190N &  10169   \\
CGCG 058-009   & IRASF 07329+1149 &0.016255 & F187N, F190N &  10169   \\
......   & IRAS 08355-4944  & 0.025898 & F130N, F132N &  13690    \\
NGC 3110   & IRASF 100150614 & 0.016858 & F187N, F190N &  10169   \\
ESO 374-IG032 & IRAS F10038-3338  & 0.03410 & F130N, F132N &  13690    \\
NGC 3256   & IRASF 102574339 & 0.009354 & F187N, F190N &  10169   \\
ESO 320-G030   & IRASF 11506-3851 & 0.010781 & F187N, F190N &  10169   \\
NGC 4922  & IRAS F12590+2934  & 0.023589 & F130N, F132N &  13690    \\
MCG 02-33-098   & IRASF 12596-1529 & 0.015921 & F187N, F190N &  10169   \\
......   & IRAS 13120-5453   & 0.030761 & F130N, F132N &  13690    \\
UGC 08387   & IRAS F13182+3424  & 0.02330   & F130N, F132N &  13690    \\
IC 0860   &  IRASF 13126+2453 & 0.011164 & F187N, F190N &  10169   \\
NGC 5135   & IRASF 13229-2934 & 0.013693 & F187N, F190N &  10169   \\
NGC 5734   &  IRASF 14423-2039 & 0.013746 & F187N, F190N &  10169   \\
NGC 5257-8   & IRAS F13373+0105 & 0.0225 & F130N, F132N &  13690    \\
NGC 5331   & IRAS F13497+0220 & 0.033043 & F130N, F132N &  13690    \\
IC 4518E/W   &  IRASF 14544-4255 & 0.015728 & F187N, F190N &  10169   \\
VV 340a    & IRAS F14547+2449 & 0.033669 & F130N, F132N &  13690    \\
ESO 099-G004  & IRAS 15206-6256   & 0.029284 & F130N, F132N &  13690    \\
NGC 5936   &  IRASF 15276+1309 & 0.013356 & F187N, F190N &  10169   \\
NGC 6090    &  IRAS F16104+5235   & 0.029304 & F130N, F132N &  13690    \\
 ......    &  IRAS F16164-0746   & 0.027152 & F130N, F132N &  13690    \\
 ......    &  IRAS F16399-0937   & 0.027012 & F130N, F132N &  13690    \\
NGC 6240    &  IRAS F16504+0228   & 0.02448  & F130N, F132N &  13690    \\
 ......   & IRASF 17138-1017& 0.017335 &  F187N, F190N &  10169   \\
IC 4687/ IC 4686   &  IRASF 18093-5744  & 0.017345 & F187N, F190N &  10169   \\
IC 4734   &  IRASF 18341-5732  & 0.015611 & F187N, F190N &  10169   \\
NGC 6701   &  IRASF 18425+6036 & 0.013226 & F187N, F190N &  10169   \\
NGC 6786    &  IRAS F19120+7320 & 0.025017  & F130N, F132N &  13690    \\
NGC 7130   &  IRASF 21453-3511 & 0.016151 & F187N, F190N &  10169   \\
IC 5179    &  IRASF 22132-3705 & 0.011415 & F187N, F190N &  10169   \\
NGC 7469   &  IRASF 23007+0836 & 0.016317 & F187N, F190N &  10169   \\
NGC 7591   &  IRASF 23157+0618 & 0.016531 & F187N, F190N &  10169   \\
 .....  & IRAS 23436+5257   & 0.034134 & F130N, F132N &  13690    \\
NGC 7771   &  IRASF 23488+1949 & 0.014267& F187N, F190N &  10169   \\
\enddata
\tablecomments{The P$\alpha$ and P$\beta$ galaxy sample. Column 1: Galaxy name; Column 2: IRAS galaxy name; Column 3: redshift; Column 4; $HST$ narrow-band filters; Column 5: $HST$ Proposal ID number}
\label{TAB:Observations}
\end{deluxetable*}
%%-------------------------------
%

%-------------------------------------------------------
\section{Data Reduction and Analysis}
%-------------------------------------------------------

The WFC3-IR images from Cycle 23 were reduced using the standard $HST$ pipeline, which includes drizzling, and stacking of the images.  
The pipeline also corrects for instrumental effects by removing bad pixels, subtracting the dark, flat-fielding and performing gain conversion. 
The Pyraf \textbf{Tweakreg} and \textbf{drizzelpac} packages were then used to align and co-add the individual exposures in each filter.   
\textbf{Tweakreg} was used to find the offsets for all images by using the position of bright stars visible in all filters. 
All of the individual images for a given object were then aligned to a common reference image using these offsets while ensuring that the co-added image for each filter would also be aligned with each other. 
Finally, the Pyraf \textbf{drizzelpac} package was used to correct for geometric distortions, remove cosmic rays, and co-add the individual images into a fully reduced mosaic image for each filter. 

\subsection{Matching physical resolution}
In order to have a consistent comparison of star-forming clumps between all the galaxies in the combined sample, we re-binned and smoothed all images to a common physical resolution of 91~pc/pixel which corresponds to the resolution of the most distant galaxy in our sample.
To create the continuum subtracted  Pa$\alpha$ and Pa$\beta$ line images, the narrow band images (F190N and F187N for Pa$\alpha$ and F132N and F130N for Pa$\beta$) were convolved with the corresponding line's point spread function (PSF) to create images with matched pixel scales and PSFs before subtraction. This allowed for a final clean continuum subtracted line image at a common pixel scale for all galaxies. Figure \ref{FIG:PaImg} shows the broad-band F110W $HST$ images as well as the continuum subtracted Pa$\beta$ line images for both NGC~6786 and NGC~6090. 
Depending on the position of the line in the narrow-band filter, there can be significant flux loss (Alonso-Herrero, 2002). We model a gaussian emission a line width of 200~km/s to estimate and correct for the flux loss due to the position of the line in the narrow-band filter. In general, the position of the Paschen line is well contained in the narrow band filter and has less than a 5\% flux loss. One exception is NGC~5257-8 which does have significant estimated flux loss of 70\% due to the position of Pa$\beta$ in the narrowband filter.

\subsection{Clump Analysis Procedure}
\begin{figure}
	\centering
	\includegraphics[width=0.49\textwidth]{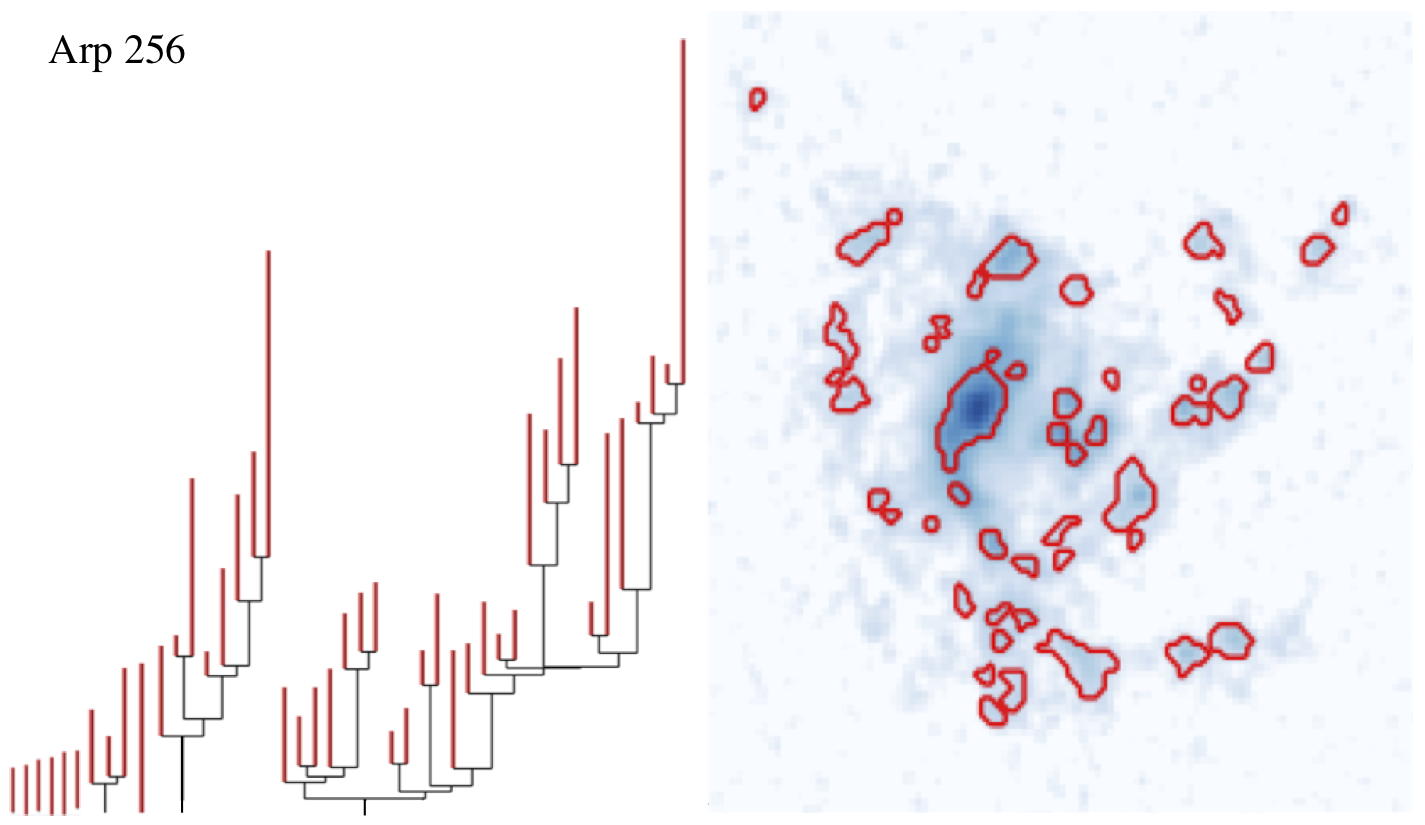}
	\caption{ \textbf{Astrodendro} is used to select individual star-forming regions in our galaxies shown here for Arp 256. Left: The dendrogram for Arp 256. The dendrogram procedure searches for all structures brighter than the local background and finds clumps down to the lowest detectable fluxes. Right: The corresponding regions for the bright structures found with \textbf{astrodendro} (shown in red) for Arp 256. } 
	\label{FIG:dendro}
\end{figure}

Our ability to compare star-forming clump properties to high-z and high-resolution simulations depends on being able to consistently detect star-forming clumps with a wide range of luminosities and sizes.
 In order to detect the clumps and measure their properties, we use the continuum-subtracted line images and the python code \textbf{astrodendro}\footnote{http://www.dendrograms.org}.
 \textbf{Astrodendro} computes a dendrogram, which is a branching tree of hierarchical structure, of an image.
 This procedure compares each possible clump to the local background, successfully identifying clumps
 down to the lowest detectable flux limit (Figure \ref{FIG:dendro}).
 %--------------------------
%--------------------------
 
 There are three important parameters in the \textbf{astrodendro} code that affect the identification of clumps. 
 First, the minimum size of the region. Since the FWHM of the data is $\sim$1 pixel, we set a minimum size of 3 pixels to be accepted as a clump. The 3 pixel minimum size maximizes the detection of the smallest clumps while avoiding blending of clumps.
 Then, a minimum threshold above the general background level is also required, which we set at 5 $\sigma$ above the general sky level of the image. This ensures that we have a resolved clump that is well above any background noise or diffuse emission. 
 Lastly, a minimum significance for the substructure is set which determines when a substructure is brighter than the local background. The minimum significance is the difference in surface density required between a structure and any substructures for a substructure to be retained. When a substructure is retained, it is shown as a new branch in the dendrogram tree in Figure \ref{FIG:dendro}. We set the minimum significance to 1 $\sigma$ above the local background. We then take the upper branch of the dendrogram tree (marked red lines in Fig. \ref{FIG:dendro} left) as our final clump regions.
 \textbf{Astrodendro} also detects any bright artifacts left in the images from bright stars or noise. Therefore, the clumps are all confirmed by visual inspection and any sources not associated with the galaxy are removed. 
  A detailed comparison of \textbf{astrodendro} to the other commonly used \textsc{clumpfind} algorithm (Williams, J. et al. 1994) for identifying and measuring clump properties in our sample LIRGs, is given in Appendix A.
 
Both the continuum emission and the line emission must be accurately measured in order to determine the ages and masses of all the clumps. The same clump regions found with astrodendro on the Pa$\alpha$ and Pa$\beta$ line images were used on the narrow-band continuum images. 
 We subtract the local background, determined by the region surrounding the clump and not including any other clumps, to measure the clump fluxes.
The local background regions were visually inspected for each clump in line and continuum images in order to avoid hotspots or other continuum features that could cause an over-subtraction of the clump line and/or continuum fluxes.
If they were contaminated by continuum features, alternative clean background regions near the source were selected. 
Local background subtraction is incredibly important to ensure that the resulting SFRs, ages, and masses are for the clump and not contaminated by underlying stellar populations (Guo et al. 2018).
We determine the errors in the clump sizes using a Monte-Carlo method. \textbf{Astrodendro} is re-run 1000 times randomly adjusting the flux of each pixel in the image while sampling from a poisson distribution. The standard deviation of the size is determined from the measured radii of each region from the 1000 iterations. This allows us to characterize the accuracy of the region's size based on the noise in the image.

%

%-------------------------------------------------------
\section{Results}
%-------------------------------------------------------

%---
 In total, we find 810 star-forming clumps: 59 nuclear clumps and 751 are extra-nuclear clumps.
The median number of extra-nuclear clumps we find per LIRG is 15.6.
The number of clumps per LIRG varies from a merging galaxy dominated by only 2 bright nuclei and no extra-nuclear clumps, to a galaxy pair with 95 extra-nuclear clumps. 
%--------------------------

\subsection{Ages and Masses}

The ages and masses of the clumps were computed by comparing the measured line and continuum fluxes of the clumps to Starburst99 models (Leitherer et al. 1999). We estimated the Pa$\alpha$ and Pa$\beta$ equivalent widths of every clump and matched them to the corresponding equivalent widths of the model spectra which were calculated in time steps of 0.1~Myr to determine the clump ages. 
We compared to both an instantaneous burst single stellar population model and a continuous star formation model with a Kroupa (Kroupa 2001) IMF and solar metallicity for our clumps. 
The masses of the clumps were calculated with the background subtracted continuum luminosities measured from the narrow band filters (F130N or F187N) and the estimated mass to light ratio from the model M(M$_\odot$) = L$_{\rm cont-filter} \times (M/L)$. The mass to light ratio varies with the age of the stellar population and was estimated for each clump at the fitted age. 

\begin{figure}[t!]
	\centering
	\includegraphics[width=0.45\textwidth]{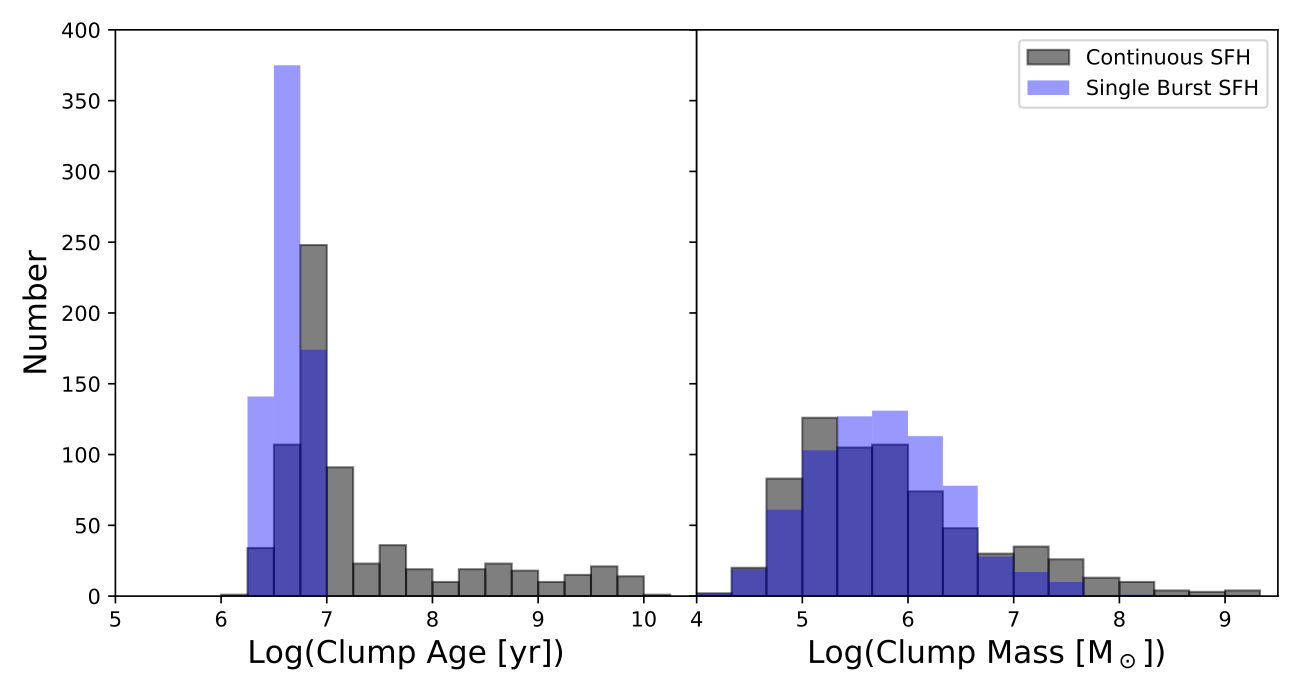}
	\caption{Distribution of the measured ages and masses for extra-nuclear star-forming clumps found in our study. The results for a continuous star formation history model are shown in black and the single burst star formation history model are shown in blue.}
	\label{FIG:AgeHist}
    \label{FIG:MassHist}
\end{figure}

Figure \ref{FIG:AgeHist} shows the distributions of ages and masses for extra-nuclear star-forming clumps.
The instantaneous burst model gave a narrow range of ages from $2 \times 10^{6}$ to $6.8 \times 10^6$~yr with a median age of 4 (+0.8, -1)$  \times 10^{6}$~yr. All uncertainties are given at the 68\% confidence levels. The continuous burst has a wider range of stellar ages from $1.5 \times 10^{6}$ to $1.6 \times 10^{10}$~yr with a median age of $8.7 (+6, -2)   \times 10^{6}$~yr.  
We found that both the instantaneous burst and continuous star formation models give masses of extra-nuclear clumps that range from $10^4$ to $10^{9}$~M$_\odot$ with a similar median clump mass of $\sim 5 \times 10^{5}$~M$_\odot$. The median and range of clump ages and masses using both the instantaneous burst and continuous star formation models for the extra-nuclear clumps is given in Table \ref{TAB:AvgProperties}. 
Since the nuclei can be contaminated by AGN and may be effected by higher levels of extinction, we exclude the nuclei from Figure \ref{FIG:AgeHist} and Table \ref{TAB:AvgProperties}.

%%%%%%%%%%%%%%%%%%%%%%%%%%%%%%%%%%%%%%%%%%%%%%%%%%%%%%%%%
%%-------------------------------
%Galaxy properties Table
\begin{deluxetable}{lllll}
\tablecolumns{6} 
\tablecaption{Average Extra-nuclear Clump Properties}
\tablehead{
 \colhead{ }  & \colhead{min} & \colhead{max} & \colhead{median} & \colhead{mean}
 }
\startdata
Log(Age)-Cont.  & 6.17 & 10.21 & 6.94 (+0.24, -0.10) & 7.33 \\
Log(Age)-Inst.   & 6.29 & 6.84  & 6.65 (+0.70, -0.11) & 6.63 \\
Log(Mass)-Cont. & 4.12 & 9.22  & 5.69 (+0.44, -0.38) & 5.87 \\
Log(Mass)-Inst.  & 4.14 & 8.31  & 5.73 (+0.36, -0.29) & 5.77 \\
SFR      & 0.002 & 4.43 & 0.03 (+0.03, -0.01)   & 0.088\\
Radius   & 89    & 678  & 171  (+47, -26) & 199  \\
$\Sigma_{SFR}$ & 0.045 & 14.86 & 0.31 (+0.18, -0.10)& 0.55 \\
\enddata
\tablecomments{The minimum, maximum, median and mean values for extra-nuclear clump properties. The 68\% confidence levels are denoted for the medians in the table. The ages and masses are given for both the continuous SFH and single burst SFH. Row 1: age Log([yr]) using the continuous SFH; Row 2: age Log([yr]) using the single burst SFH; Row 3: mass Log([M$_\odot$]) using continuous SFH; Row 4: mass Log([M$_\odot$]) using single burst SFH; Row 5: SFR [M$_\odot$ yr$^{-1}$]; Row 6: effective radius [pc], the minimum radius is set by the minimum allowed clump size of 3 pixels; Row 7: SFR surface density [M$_\odot$ yr$^{-1}$ kpc$^{-2}$] }
\label{TAB:AvgProperties}
\end{deluxetable}
%%%%%%%%%%%%%%%%%%%%%%%%%%%%%%%%%%%%%%%%%%%%%%%%%%%%%%%%%

\begin{figure}[t!]
	\centering
	\includegraphics[width=0.45\textwidth]{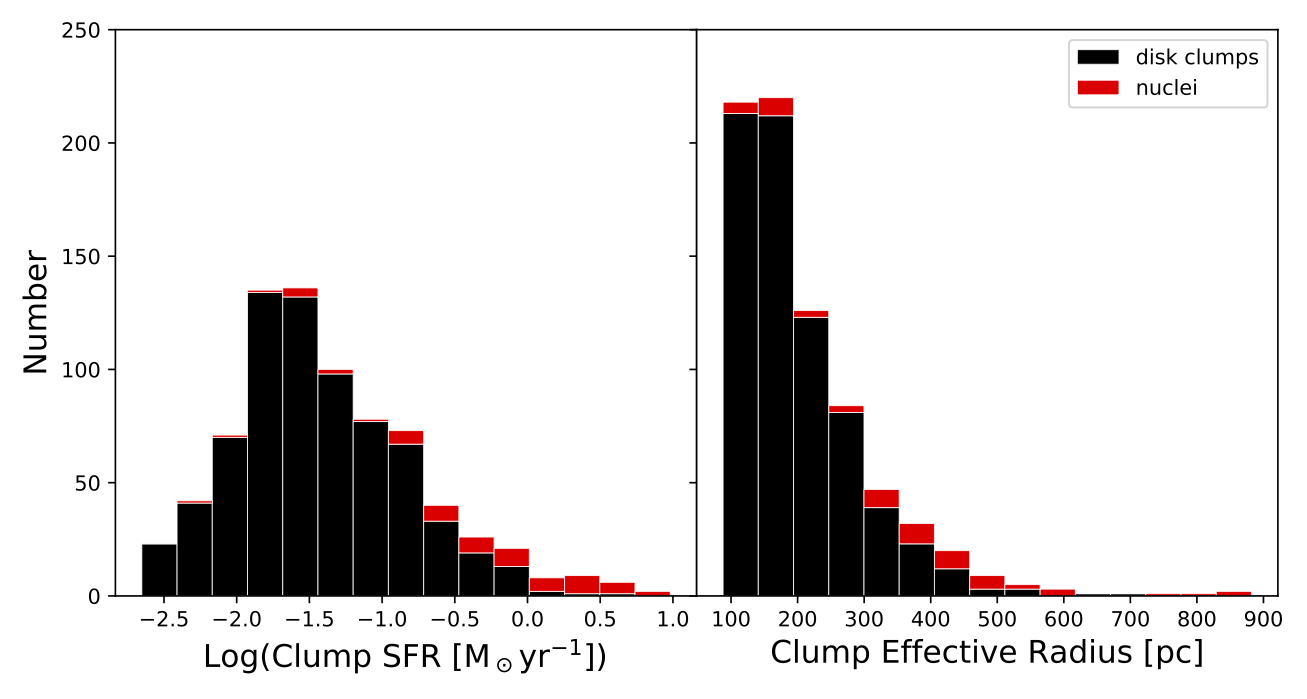}
	\caption{Distribution of the measured SFR and sizes for all regions. Extra-nuclear clumps are shown in black while nuclear clumps are red. The Paschen line luminosities were converted to SFRs based on a continuous star formation history model.}
	\label{FIG:SFRSizehist}
\end{figure}

% clumps with uncertain ages and masses
Even with careful background subtraction, we were unable to calculate accurate ages and masses for 45 extra-nuclear clumps. These clumps have large errors on the continuum measurements which lead to errors in the age estimate that are as large as the age. They have very little continuum detected and therefore large errors that do not allow for accurate estimates of the ages and masses. We therefore exclude these clumps from our discussion of ages, masses, and SFRs. Further discussion of ages, masses, and SFRs of extra-nuclear clumps will only include the 706 clumps.
A further 21 of these extra-nuclear clumps have unconstrained ages due to equivalent widths that are outside of the range predicted by the continuous SFH model. These clumps are clearly identified in both line and continuum emission and are therefore included in the future discussion of SFRs.

\subsection{Star Formation Rates and Sizes}

The clumps we find using the astrodendro technique span a large range of luminosities and sizes as shown in Figure 4. 
The area of the clump is determined by the number of pixels in the region and the effective radius is defined as r= $\sqrt{A/\pi}$.
To allow for an equal comparison of both Pa$\alpha$ and Pa$\beta$ images and other H$\alpha$ clump studies, we convert all clump luminosities to SFRs. We use a conversion factor determined from Starburst99 stellar population models assuming a continuous star formation, Kroupa IMF, and stellar metallicity at the appropriate age for each clump. 
This conversion factor uses the `Case B' line intensity ratios and assumes no extinction. 
While extinction from dust obscuration is much less in the NIR than in the optical for U/LIRGs, it can still affect the nuclei of the galaxies and decreases quickly in the extra-nuclear regions (D\'iaz-Santos et al., 2011). 
Previous studies using VLT-SINFONI data have shown that local (z$<$0.02) LIRGs have sub-kiloparsec clumpy dust structures. These regions have a wide range of A$_{\rm{v}}$ from 1$\sim$20mag and an average A$_{\rm{v}}$=5.3 mag within their 3x3~kpc FoV (Piqueras Lopez et al, 2013). We therefore can not apply a single galactic A$_{\rm{v}}$ to our star-forming clumps since the dust obscuration is extremely patchy and each galaxy presents different spatial clump distributions. In a study of star clusters in GOALS galaxies, Linden et al. (2017) estimated the extinction to individual extra-nuclear clusters using FUV, B and I photometry. Thirty-two of our star-forming clumps distributed over 6 galaxies overlap with their sample of detected extra-nuclear star clusters. The estimated A$_{\rm{v}}$ for these extra-nuclear regions ranges from $\sim 0.2 - 1.8$~mag with an average A$_{\rm{v}}$ = 1, which implies a A$_{\rm{NIR}}$ $\sim 0.1-0.2$~mag. Therefore while extinction corrections can be important for some extremely dusty clumps, the extinction is highly spatially variable and can be fairly low outside of the circum-nuclear regions.  Our nuclear clumps likely have the largest corrections, but these are unknown.  The fluxes, and derived star formation rates of the nuclear clumps are therefore, effectively, lower limits.

Due to large continuum measurements and accordingly low equivalent width in some of the nuclei, 20 of the 59 nuclear clumps have equivalent widths that are below the range predicted by the continuous SFH models. For the 20 nuclei and the 21 extra-nuclear clumps with equivalent widths not covered in the continuous SFH models, we cannot determine an age.   The clump SFRs are therefore calculated using the asymptotic calibration value
from the Starburst99 continuous SF model at $10^8$~yrs (SFR/L$_{H\alpha} = 5.37 \times 10^{-42}$ [M$_\odot$yr$^{-1}$]/[erg~s$^{-1}$]).
While the Starburst99 continuous SF model provides a variable luminosity to SFR calibration with age, the calibration starts to asymptote to a constant value at an age of $5\times10^6$~yrs and reaches a constant value by $10^8$~yrs.
The asymptotic calibration reached for our Starburst99 models is the same calibration determined by Murphy et al. (2011). 
%The Murphy et al. (2011) calibration is determined using the same Starburst99 continuous SF model assuming an age $>$10Myr.

\begin{figure}[t!]
	\centering
	\includegraphics[width=0.49\textwidth]{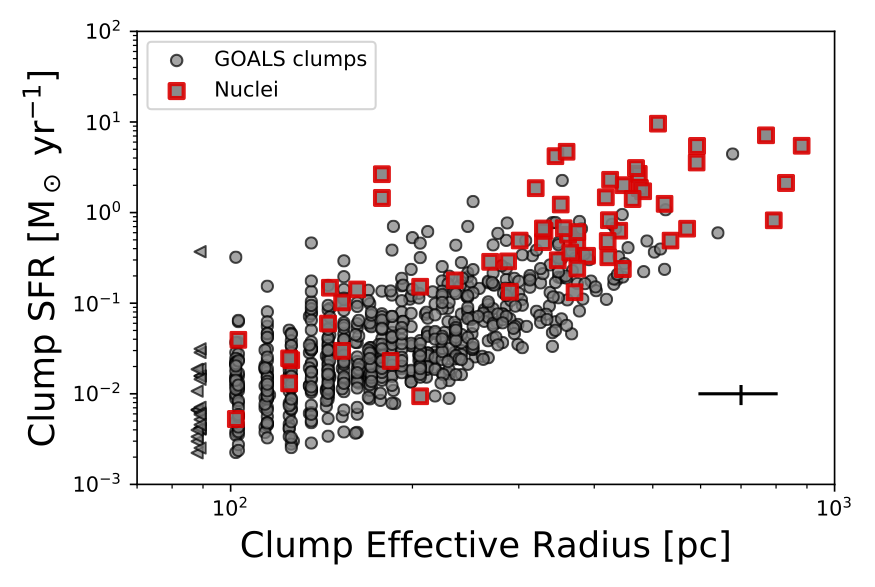}
	\caption{The individual clump SFRs as a function of size for all star-forming regions found in the GOALS Pa$\alpha$ and Pa$\beta$ sample. Some of the brightest and largest clumps coincide with the nuclei of the galaxies, and these are marked with a red box. \textcolor{black}{The sizes of the smallest clumps are upper limits and are indicated with a left triangle. Average error bars for the data are given in the lower right corner.}   }
	\label{FIG:SFR_size}
\end{figure}

In Figures \ref{FIG:SFRSizehist} and \ref{FIG:SFR_size}, we see that the clump SFRs span a wide range from $2\times10^{-3}$ to 4.4~M$_{\odot}$yr$^{-1}$  with nuclear clumps that have SFRs of $5\times10^{-3}$ to $9.6$~M$_{\odot}$yr$^{-1}$.
We find clump radii that range from our minimum resolvable size of 89~pc up to 883~pc with extra-nuclear clumps reaching sizes of 678~pc and a median of 171 (+47, -26)~pc,
%%%%%%%%%%%%%%%%%%%%%%%%%%%%%%%
%%%%%%%%%%%%%%%%%%%%%%%%%%%%%%
We provide the measurements of all clumps with determined SFRs in our Pa$\alpha$ and Pa$\beta$ sample in Table \ref{TAB:ClumpProp1}. The clumps are identified by their host galaxy and ID number (GalaxyName\textunderscore c\#). Data for nuclei in each galaxy are also given and identified by the host galaxy name and nuclei number (GalaxyName\textunderscore n\#).   The estimated age and mass of each clump is given, assuming a continuous star formation history and solar metallicity, as well as the clump SFR and effective radius. 

%%%%%%%%%%%%%%%%%%%%%%%%%%%%%%%%%%%%%%%%%%%%%%%%%%%%%%%%%
\begin{table*}[t]
	\caption{Properties of star-forming clumps in GOALS galaxies. This table is available in its entirety in a machine-readable format online.}
	\begin{center}
	\begin{tabular}{|l|l|l|l|l|l|l|l|l|}
	\hline
 Clump   & Radius& err &  SFR & err &  Age &err &   Mass &err  \\
 Name     & [pc]& [pc]  &  [M$_\odot$ yr$^{-1}$]&  [M$_\odot$ yr$^{-1}$]  & [yr]& [yr]  & [M$_\odot$] & [M$_\odot$]   \\

\hline
MCG+02-20-003\textunderscore n1    & 269 & 24 & 0.286 & 0.009 & NaN &NaN & NaN & NaN   \\
MCG+02-20-003\textunderscore c1    & 254  & 80 & 0.026 & 0.007 & 6.1e+08 & 4.5e+08 & 1.5e+07 & 1.2e+07     \\
MCG+02-20-003\textunderscore c2    & 103 &21 & 0.006 & 0.001  & 3.6e+09 & 1.5e+09 & 1.8e+07 & 6.7e+06    \\
MCG+02-20-003\textunderscore c3    & 187 & 29 & 0.008 & 0.002 & 3.8e+09 & 1.8e+09 & 2.6e+07 & 1.2e+07   \\
MCG+02-20-003\textunderscore c4    & 127 & 24&  0.003 & 0.001  & 1.8e+09 & 1.1e+09 &  4.9e+06 & 2.6e+06  \\
MCG+02-20-003\textunderscore c5    & 116 &22&  0.010 & 0.001 & 1.3e+07 & 3.0e+06 &  1.2e+05 & 4.8e+04  \\
MCG+02-20-003\textunderscore c6    & 226 &19&  0.048 & 0.005 & 1.0e+07 & 2.0e+06 &  4.9e+05 & 2.8e+05  \\

	\hline
	\end{tabular}
\end{center}
        \begin{tablenotes}
        \small
        \item[]    \textbf{NOTE--} Properties of star-forming clumps found in GOALS galaxies. Column 1: Source name; Column 2: effective radius of the clump in pc calculated from the clump area. Upper limits are indicated with a negative value; \textcolor{black}{Column 3: radius error [pc]}; Column 4: clump SFR in M$_\odot$~yr$^{-1}$ calculated from the Pa$\alpha$ or Pa$\beta$ flux; Column 5: clump SFR error; Column 6: clump age [yr] calculated from the estimated equivalent width; Column 7: clump age error [yr]; Column 8: clump mass in M$_\odot$; Column 9: clump mass error %calculated from the narrow-band continuum 
        \end{tablenotes}
\label{TAB:ClumpProp1}
\end{table*}

%%%%%%%%%%%%%%%%%%%%%%%%%%%%%%%%%%%%%%%%%%%%%%%%%%%%%%%%%

\begin{figure}[t!]
	\centering
	\includegraphics[width=0.49\textwidth]{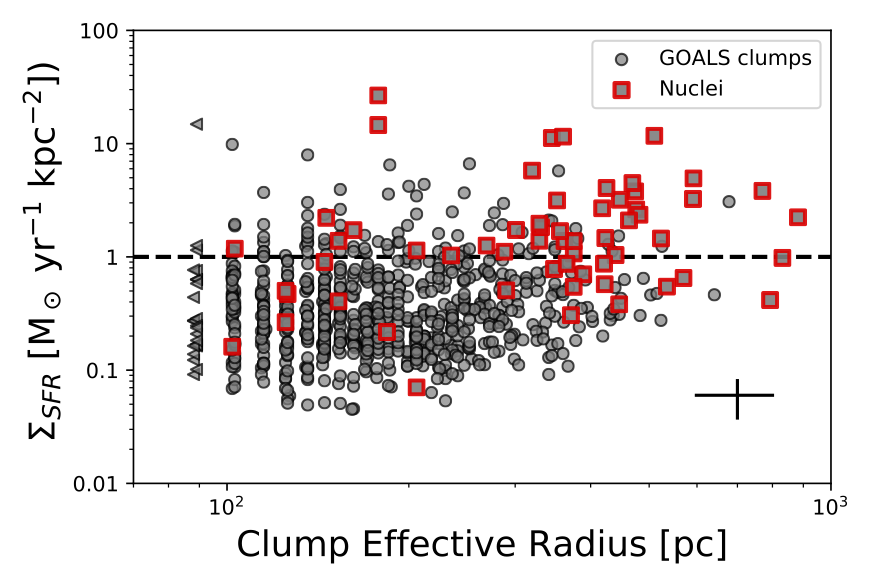}
	\caption{The star formation rate density of the Pa$\alpha$ and  Pa$\beta$ clumps as a function of clump size. The nuclear clumps are marked with a red box and, as expected, they typically have larger sizes and higher luminosities than the star-forming clumps in the disks. \textcolor{black}{The sizes of the smallest clumps are upper limits and are indicated with a left triangle. Average error bars for the data are given in the lower right corner.} }
	\label{FIG:SFRDens_size}
\end{figure}

The star formation rate surface density ($\Sigma_{SFR}$) of the extra-nuclear GOALS clumps range from 0.045 to 14.86~M$_\odot$yr$^{-1}$kpc$^{-2}$. Star-forming regions with  $\Sigma_{SFR} > 1$~M$_\odot$yr$^{-1}$kpc$^{-2}$ are often referred to as `starbursting' since extra-nuclear starforming regions in normal spiral galaxies are found to have $\Sigma_{SFR}$ below this cutoff (Maragkoudakis et al. 2016; Nguyen-Luong et al. 2016; Lacki et al. 2013; Alonso-Herrero et al. 2000; Kennicutt 1998). Figure \ref{FIG:SFRDens_size} shows how the majority of the clumps lie below $\Sigma_{SFR} = 1$~M$_\odot$yr$^{-1}$kpc$^{-2}$ with 13\% of the extra-nuclear clumps in the `starburst' region. The median $\Sigma_{SFR}$ of extra-nuclear clumps is 0.31 (+0.19 ,-0.10)~M$_\odot$yr$^{-1}$kpc$^{-2}$ while nuclear clumps are have a higher median $\Sigma_{SFR} = $ 1.4 (+0.8, -0.5)~M$_\odot$yr$^{-1}$kpc$^{-2}$ and are more likely to be starbursting with 64\% of the nuclear clumps in the `starburst' region. The median and range of clump SFRs and $\Sigma_{SFR}$s for the extra-nuclear clumps is given in Table \ref{TAB:AvgProperties}.
Piqueras L\'opez et al. (2016) 
observed a sample of 17 LIRGS and ULIRGS with the VLT/SINFONI integral field unit and found a wider range of star formation rate surface densities for clumps from $0.1 - 60$ $\Sigma_{SFR}$, although these data were corrected for extinction (from the Br$\gamma$ and Br$\delta$ ratios). 
\textcolor{black}{We converted the Piqueras L\'opez et al. (2016)  results from Salpeter to Kroupa IMF to be comparable to the results presented in this paper. The wide range of star formation rate surface densities found by Piqueras L\'opez et al. (2016) is due to their applied extinction correction. The clumps with the largest $\Sigma_{SFR}$ in their study are the ones with the largest inferred extinction, and these tend to be in the nuclei, making up a small fraction of the distribution.}
Therefore, the median observed and extinction corrected $\Sigma_{SFR}$ for the ULIRGs in the Piqueras L\'opez et al. (2016) sample are $\Sigma_{SFR}^{obs}= $0.11~M$_\odot$yr$^{-1}$kpc$^{-2}$ and 
$\Sigma_{SFR}^{corr}= $0.16~M$_\odot$yr$^{-1}$kpc$^{-2}$, similar to the median extra-nuclear $\Sigma_{SFR}$ found in our sample of $\Sigma_{SFR}=0.31$ . 

Even though our galaxy sample is selected to be infrared luminous, the majority of our extra-nuclear star-forming clumps have $\Sigma_{SFR}$ at the upper end of $\Sigma_{SFR}$s that are found in normal spiral galaxies with 13\% of the clumps in the starburst region.

%%%%%%

\begin{figure*}[t!]
    \includegraphics[width=0.49\textwidth]{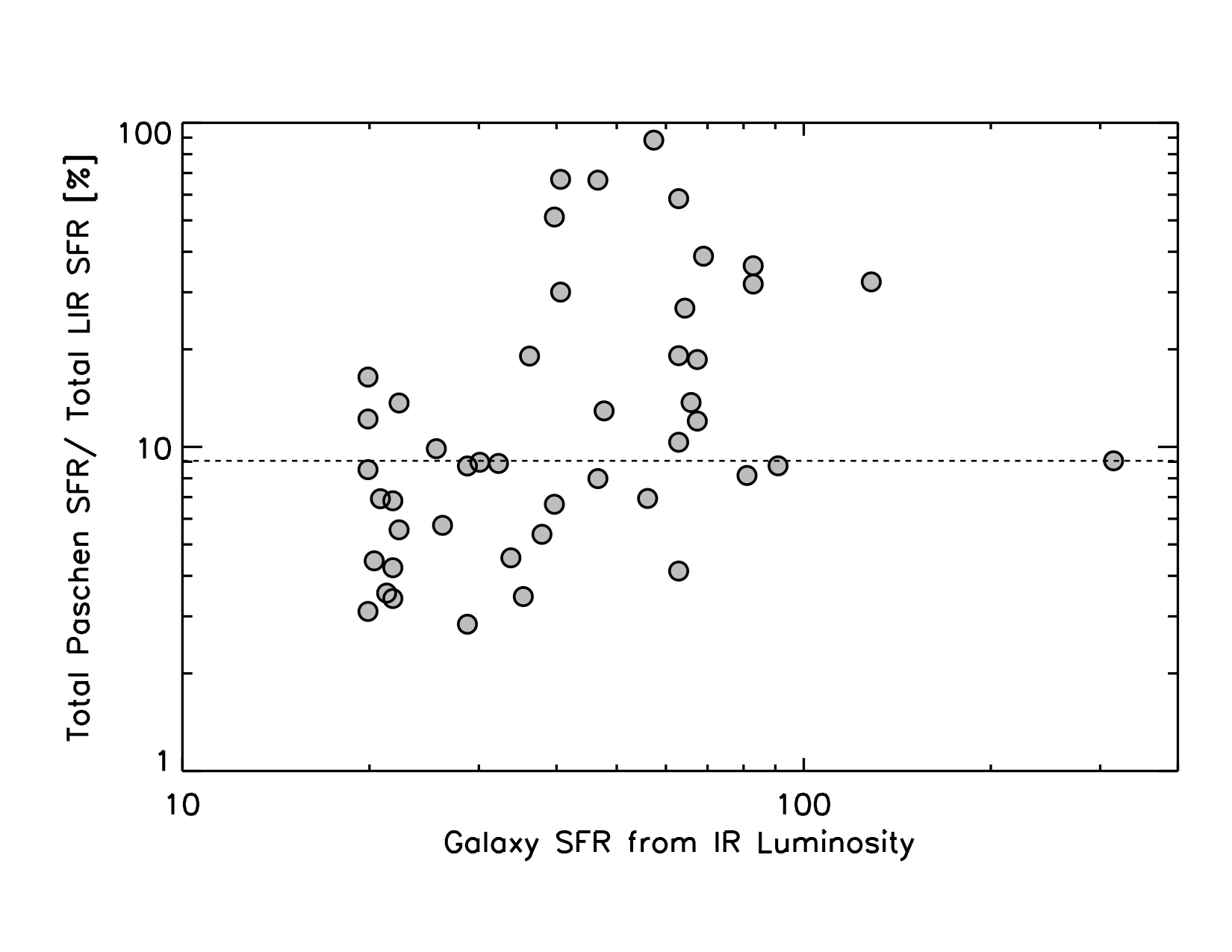}
    \includegraphics[width=0.49\textwidth]{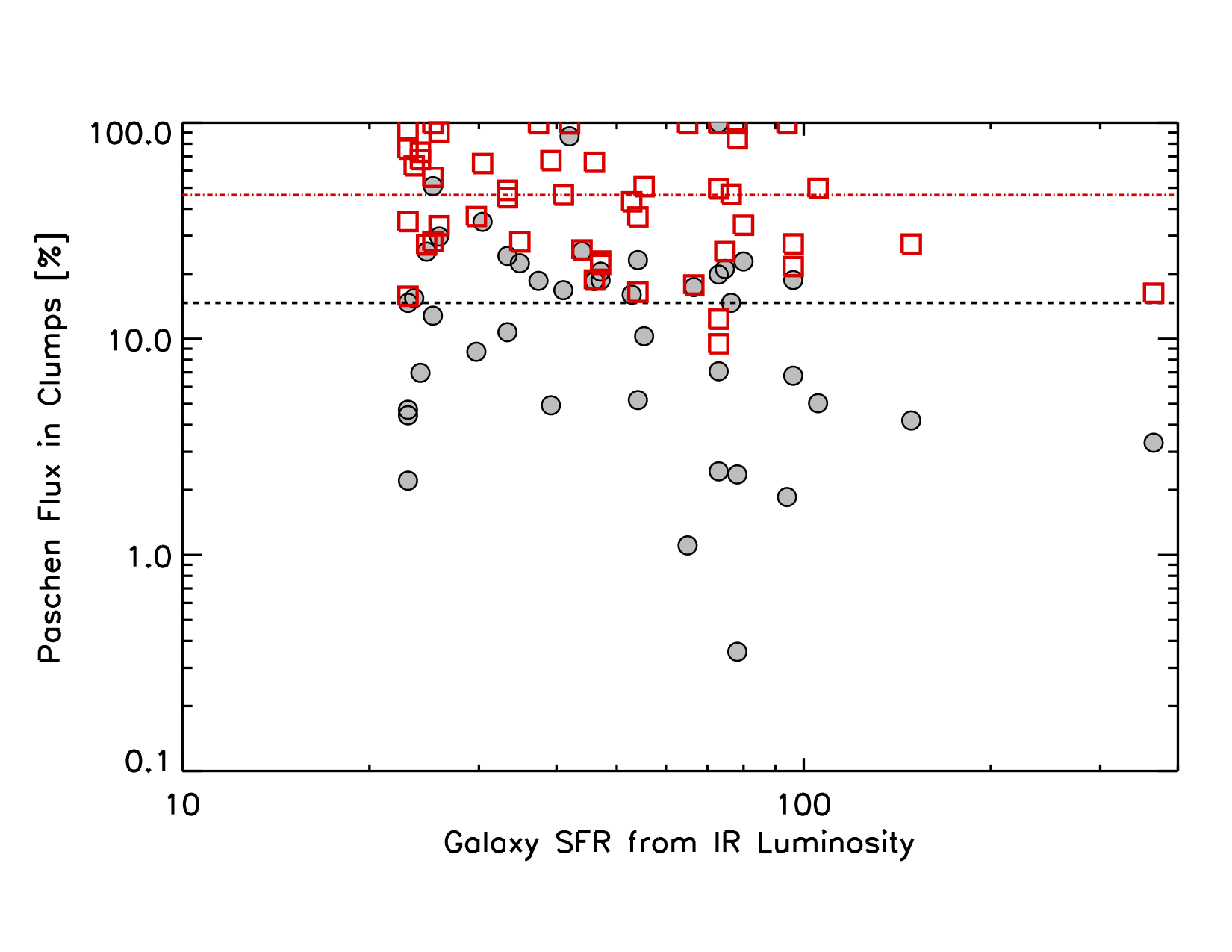}
	\caption{Left: The percentage of the total far-infrared-derived SFR of the galaxies observed in Pa$\alpha$ or Pa$\beta$ (including clump and diffuse emission) as a function of the far-infrared-derived total SFR. Each point represents the percentage for an entire galaxy. A median of 9\% of the total far-infrared-derived SFR of the galaxy is recovered in the Pa$\alpha$ or Pa$\beta$ lines and is shown as a dotted line. \textcolor{black}{One galaxy, IC 0860, has only 0.14\% of the total infrared emission detected and is therefore not included in the figure.}
	Right: The percentage of the total Paschen flux of the galaxies observed in extra-nuclear star-forming clumps are plotted as a black plus symbol and the percentage of flux in clumps including the nuclei are plotted as blue squares. Each point represents the percentage for an entire galaxy. The median value of 14.7\% for the non-nuclear clumps is shown as a black dashed line. Including the SF from the nuclei increases the median SF in the clumps to 46.3\% and is shown as a blue dot-dash line.}
    \label{FIG:PercClumps}

\end{figure*}

The total star formation rate, derived from the $HST$ data, includes the clumpy and diffuse narrow-band emission. For each galaxy, this is determined by converting the Paschen luminosity to H$\alpha$ luminosity assuming the `Case B' recombination, and then converting the H$\alpha$ luminosity to SFR using the calibration from Murphy et al. (2011)
(SFR/L$_{H\alpha} = 5.37 \times 10^{-42}$~[M$_\odot$yr$^-1$]/[erg~s$^{-1}$]). This calibration assumes a Kroupa IMF with continuous SFR and solar metallicity integrated over a timescale of 100 Myr. We use the Murphy (2011) calibration for obtaining the total SFR of the galaxies since we are averaging over a range of sizes and ages of star-forming clumps and diffuse emission.
Figure \ref{FIG:PercClumps} shows the fraction of the total star formation rate, as derived from the infrared emission, is detected in each galaxy in the narrow-band $HST$ images.
%convert ir lum to sfr... 
The total infrared SFR of the galaxies is calculated by converting the total infrared luminosity of the galaxies, L$_{IR}$, to a SFR (Murphy et al. 2011).

%---------------------------------
%---------------------------------

There is a large range in the percentage of the total star formation rate, as traced in the far-infrared, recovered in our Paschen data (Fig.7).  The median ratio of P$\beta$ or P$\alpha$ to total far-IR derived star formation in our sample is 9\%.
The galaxy with the lowest detected Paschen emission is IC 0860, with only 0.14\% of the total infrared emission detected and is not shown in Figure \ref{FIG:PercClumps}. This galaxy has no extra-nuclear clumps and all detected emission is from the nucleus.  This source must be heavily enshrouded even in the near-infrared.

Figure \ref{FIG:PercClumps} shows the percentage of the
Paschen emission for each galaxy observed in clumps. The percentage of the total Paschen emission contained in extra-nuclear clumps is 0 to 65\% with a median of 14.7\%. When the nuclei are included, 10 to 100\% of the Paschen emission is contained in clumps with the remaining emission in diffuse emission.
The nuclear regions can contain a significant fraction of the detected Paschen emission even without extinction correction.
Some galaxies -like NGC 5258 and NGC 6786- are dominated by extra-nuclear star formation with no detectable nuclear emission in Pa$\beta$. While other galaxies -like IRAS F02437+2122, IRAS 23436+5257, and NGC 7591- are dominated by strong nuclear emission with little or no extra-nuclear star-forming clumps.

%------------------
\section{Discussion}
%------------------

\subsection{Luminosity Function}

The luminosity function (LF) of star-forming clumps in local galaxies can be well approximated by a power law with a slope of $\alpha=-2$ (e.g. Zhang \& Fall 1999; de Grijs et al. 2003; Chandar et al. 2015, Cook et al. 2016). 
Local LIRGs cover SFR of 20$<$ SFR~[M$_\odot$yr$^{-1}$] $<$200, thus making up the high LF end.
\begin{figure*}[t!]
    \includegraphics[width=0.49\textwidth]{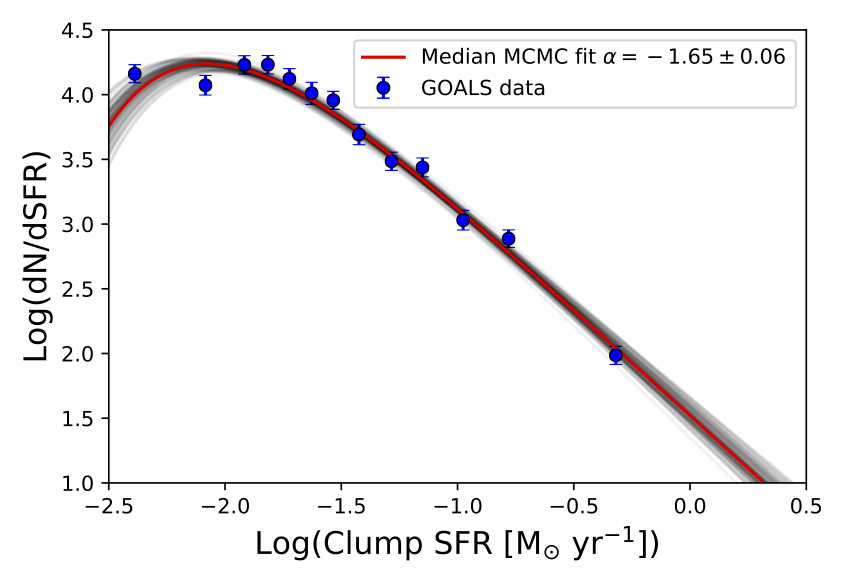}  \includegraphics[width=0.49\textwidth]{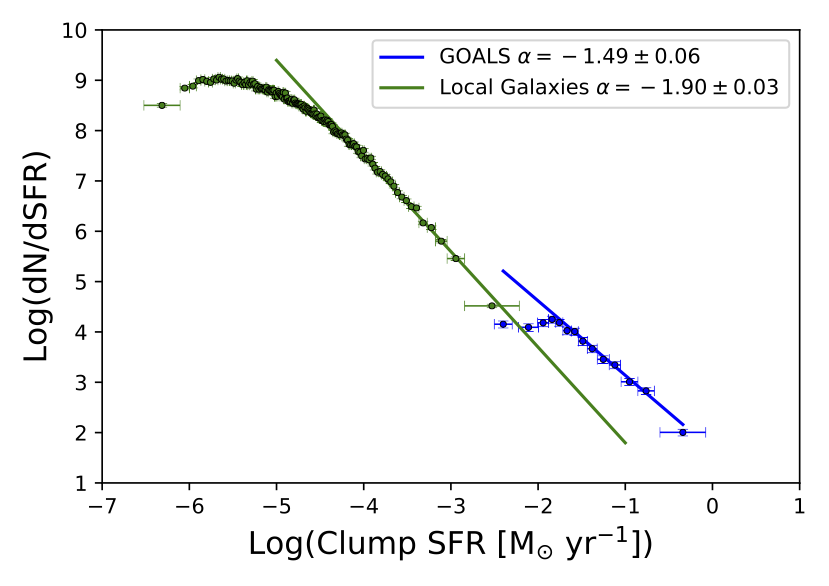}
    \caption{Left: Composite star formation rate function for GOALS extra-nuclear star-forming regions. The function is modeled by a single power law plus a low-SFR exponential cutoff which gives a slope of $\alpha = -1.65 \pm 0.06$ and a log(SFR) cut-off $= -1.82 (+.04, -.05)$~M$_\odot$yr$^{-1}$. A random selection of MCMC-derived SFR functions are overplotted in grey. 
	Right: Composite SFR functions for UV star-forming regions from the galaxies in Cook et al. (2016), shown in green, and the extra-nuclear star-forming regions from GOALS galaxies, in blue. Both functions are fit at the bright end where they are most complete. The GOALS sample shows a shallower slope ($-1.49 \pm 0.06$) than local, normal, non-merging galaxies ($-1.90 \pm 0.03$). This may be due to the presence of larger numbers of bright, dense clumps in the GOALS galaxies.  }
    \label{FIG:MCMCLumF}
     \label{FIG:LumF}
\end{figure*}

 Previous studies have shown that different binning methods can change the measured slope of the LF (Ma\'iz, Apell\'aniz \& \'Ubeda 2005, Cook et al. 2015). While the equal luminosity-sized bins is the most commonly used method, Ma\'iz, Apell\'aniz \& \'Ubeda (2005) showed that this caused systematic biases at the brightest luminosity bins due to low number statistics. Therefore, we chose to use variable bin sizes where each bin contains an equal number of sources.
 
While we do not have enough clumps per galaxy to do individual LFs for each galaxy in our sample, we can create a combined function of clumps from all GOALS galaxies. 
Since we are comparing data from a combined P$\alpha$ and P$\beta$ sample, we create a star formation rate function using the SFRs of the clumps as defined previously in section 3.3.
Only extra-nuclear clumps are included in the star formation rate function.

We fit the data in two ways to determine the slope.
First, we determine the slope by MCMC modeling of the star formation rate function as a power law plus an exponential cutoff. This simultaneously fits the slope and the incompleteness cutoff point giving a slope of $\alpha = -1.65 \pm 0.06$ and a SFR cut-off $= 0.015 \pm 0.002$~M$_\odot$yr$^{-1}$ as shown in Figure \ref{FIG:MCMCLumF}(Left). The flattening of the number of sources at lower SFRs is due to an incompleteness in the data at the faint end. 

Second, we fit a single power law to only the bright end of the SFR function, where the data are complete. We also bin the data using an equal number per bin and determine the error on each bin by bootstrap re-sampling so we can compare directly to the results from Cook et al.(2016). 
The SFR cutoff for the fit is determined by the peak of the SFR histogram, at log(SFR cutoff)$=-1.7$. 
We find that the combined SFR function of clumps from all GOALS galaxies has a bright end slope of $\alpha=1.49 \pm 0.06$.

A study of UV star-forming regions in nearby galaxies by Cook et al. (2016) investigated the dependence of the clump LF on the galaxy star formation rates ($10^{-3}<$ SFR~[M$_\odot$yr$^{-1}$] $<$3). They calculated the SFRs of clumps in their sample by converting dusted-corrected FUV fluxes to SFRs using the Murphy et al. (2011) prescription and Kroupa IMF. 
Cook et al. (2016) found that galaxies with higher SFRs tend to have flatter clump LF slopes. They then derived a combined LF of clumps from 134 galaxies and found a bright-end slope of $\alpha= -1.93$. The galaxies in the Cook sample range in distance from $\sim$1~Mpc to $\sim$10.5~Mpc, corresponding to regions with a diameters of $\sim$24 -- 250~pc.
In figure \ref{FIG:LumF}, we also fit only the local normal galaxies having distances larger than 5.8~Mpc (equivalent to a lower limit on the clump resolution of about 70~pc) to match the GOALS sample, and find a bright end slope of $\alpha= -1.90 \pm 0.03$, still within the limits of what was initially found with the full sample. 
This demonstrates that the flattening of the slope that we see in the GOALS sample is not solely due to resolution effects.

The combined SFR function slope from the normal galaxies is steeper than that found in the GOALS galaxies as can be seen in Figure \ref{FIG:LumF}.
While both studies suffer from incompleteness at the faint end, it is clear that the bright end slopes are significantly different, suggesting an overabundance of bright, SF extra-nuclear clumps in LIRGs. 
Furthermore, the variable nature of extinction in LIRGs and ULIRGs means that some of the clumps in our sample could have Av as high as 10-20 even if on average the Av is much lower (Piqueras Lopez et al, 2013, 2016). While we can not correct for extinction directly in our data, if there was a significant effect, it would imply many more luminous clumps, and further flatten the slope of the luminosity function increasing the difference between the local, normal galaxies and GOALS galaxies.

%--------------------------------------------
\subsection{Luminosity and Size}
%--------------------------------------------

\begin{figure*}[h!]
	\centering
	\includegraphics[width=0.9\textwidth]{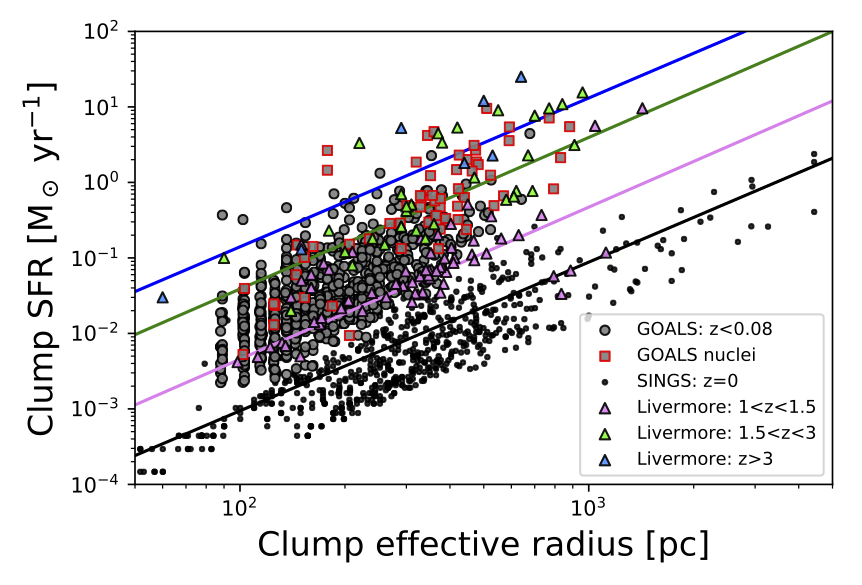}
	\caption{The clump SFR as a function of size for GOALS galaxies (gray) for both nuclear and extra-nuclear clumps, where the nuclear clumps are marked with a red box, compared to clumps from Livermore's 2012, 2015 lensed sample (triangles) and local z=0 SINGS galaxies (black circles). Livermore's lensed clump sample is divided into three redshift bins of $1.0<$ z $<1.5$ in purple, $1.5<$ z $<3$ in green, and z$>3$ in blue. The lines are constant surface brightness fits to the four redshift bins not including the GOALS SF clumps. Livermore found clumps of higher surface brightness for galaxies at higher redshift.}
	\label{FIG:GOALS_Liv}
\end{figure*}

While the range of sizes of star-forming clumps in local LIRGs are similar to those in local, normal (L$_{\rm IR} < 10^{11}$) galaxies, the LIRG clumps have much larger star formation rates (see Figure \ref{FIG:GOALS_Liv}).  Many are comparable to star-forming clumps seen in high redshift galaxies.
A significant fraction of high-z galaxies (1$<$ z $<$3) have turbulent, clumpy, disks with clump masses of $\sim 10^{8-9}$~M$_\odot$ and sizes of 0.5 -- 5~kpc (e.g. Elmegreen et al. 2004, 2009, Daddi 2010). Studying UV star-forming clumps in galaxies in the Hubble Ultra Deep Field at redshifts between 0.5$<$z$<$1.5 Soto et al. (2017) found an average clump SFR of 0.014~M$_\odot$yr$^{-1}$ and an average clump radius of 0.9~kpc with some extending up to 2~kpc. The current interpretation is that these large clumps with high SFR have higher gas fractions and increased star formation efficiency (Tacconi et al. 2008, Jones et al. 2010).

   However, most high-redshift clump studies can only reach down to resolutions of $\sim0.5-1$~kpc, which is larger than most of the resolved clumps in the GOALS sample. Lensed galaxies provide a way to
   study high-redshift, clumpy star formation on scales of $\sim$100~pc (Livermore et al. 2012, 2015). We can therefore directly compare our GOALS star-forming clumps to those found in the lensed galaxy samples.

The SFRs of GOALS clumps (10$^{-3}$--10 [M$_\odot$yr$^{-1}$] ) span the range from SINGS z=0 galaxies to those found in z=1--3 lensed galaxies; thus, bridging the gap between the local universe and the high-z universe when comparing samples at similar resolutions.
While local LIRGs have smaller fractions of high luminosity clumps than galaxies at z $>$ 2, they provide compelling evidence that the physical conditions driving extreme star formation at high-z, can be found in merging starburst galaxies at low-redshift.
We note that clumps measured in high-redshift galaxies without the benefit of lensing, would lie in the upper right corner of Figure \ref{FIG:GOALS_Liv} with sizes from 1--5~kpc and SFRs that range from $\sim$1--30~M$_\odot$yr$^{-1}$ (Genzel et al. 2011, Swinbank et al. 2012, Wisnioski et al. 2012, Soto et al. 2017, Guo et al. 2018).

%------------------------------------------
\subsection{Star-forming clumps in high-resolution cosmological hydrodynamical simulations}
%------------------------------------------
Simulations of high-z systems have recently been used to predict the distribution in sizes, luminosities and lifetimes of the gas clumps in terms of the evolution of gas-rich disks at $z \sim 2$ (Oklopcic et al. 2016).  It is natural, therefore, to compare our GOALS results to these models to see if they are similar and, if so, whether or not we might gain a better understanding of the nature and fate of the luminous star-forming clumps we see in GOALS.

\begin{figure}[!t]
	\centering
	\includegraphics[width=0.49\textwidth]{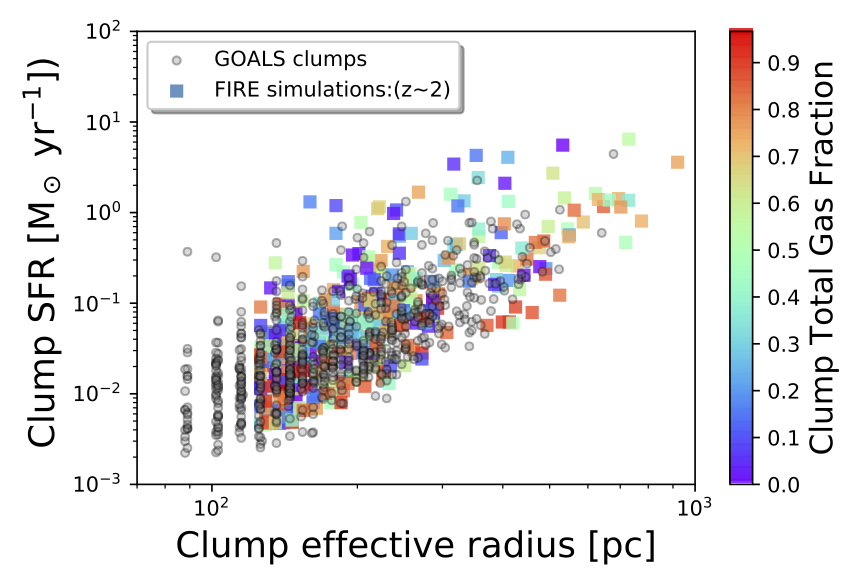}
	\caption{Comparison of extra-nuclear GOALS SF clumps (grey circles) and star-forming clumps identified in MassiveFIRE simulation (colored squares).The color of the MassiveFIRE symbols corresponds to the estimated total (atomic plus molecular) gas fraction of the clump. Blue symbols have gas fractions of $\sim$0.10 and red symbols have gas fractions of $\sim$0.90.}
	\label{FIG:FIRE}
\end{figure}

The MassiveFIRE (Feedback in Realistic Environments) simulation has 8 times smaller mass resolution ($m=3.3\times 10^{-4}$~M$_{\odot}$) than FIRE and simulates galaxies at $z > 1.7$ (Feldmann et al. 2016). The high mass resolution obtained in MassiveFIRE resolves down to a SFR density of 0.07~M$_\odot$yr$^{-1}$kpc$^{-2}$ and allows for a direct comparison to the observed star-forming clumps in GOALS. We have used the same software suite (\textbf{astrodendro}) to identify star-forming clumps in the MassiveFIRE simulation.
Using \textbf{astrodendro} on the MassiveFIRE simulations,
resolves SF clumps down to a size of 126~pc effective radius allowing for a relevant comparison to our GOALS SF clumps. Figure \ref{FIG:FIRE} shows the SFR versus size for both the GOALS and MassiveFIRE star-forming clumps. The clumps in MassiveFIRE have a similar range of SFRs (4$\times10^{-3} - 6.45$~M$_\odot$yr$^{-1}$) and sizes (126~pc -- 1~kpc) to the GOALS star-forming clumps. Since the FIRE simulations model both the stars and gas, we can use these results to 
estimate basic properties (e.g. the star formation efficiencies) of the star-forming clumps in GOALS. 

The models predict star-forming clumps with a wide range of total (atomic plus molecular) gas fractions from $\sim$ 10\% to $\sim$ 90\% gas. Higher star formation clumps have lower gas fractions. Most of the GOALS clumps that overlap in SFR-size space with MassiveFIRE clumps have intermediate to high clump gas fractions ( $> 50$\%), while clumps with the highest star formation rates, and lowest gas fractions, do appear rare in the GOALS data.
The MassiveFIRE clumps have a median star formation efficiency of $3.5\times10^{-9}$~yr$^{-1}$, almost an order of magnitude higher than the kpc scale star formation efficiency found in local spiral galaxies of $4.3\times10^{-10}$~yr$^{-1}$ (Leroy et al. 2008). Similarly high star formation efficiency has been observed for individual 250~pc regions in one galaxy from our sample, IC~4687,  which has both ALMA CO(2-1) and HST Pa$\alpha$ data (Pereira-Santaella et al. 2016).

A pilot study with ALMA is currently underway to directly measure the resolved molecular gas content on sizes comparable to our star-forming clumps. These ALMA observations of the GOALS sources will allow us to test for possible correlations with gas content, and directly measure the star formation efficiencies of these clumps. In our next paper we will also investigate if galaxy properties, like the merger stage, are driving the SFRs in our clumps.

%--------------------------------------------
%--------------------------------------------

%--------------------------------------------
\section{Conclusions}
%--------------------------------------------

 We use narrow-band $HST$ imaging to target Pa$\alpha$ and Pa$\beta$ emission and trace the star formation in 48 local luminous infrared galaxies. The LIRGs cover a range of merger stages from single isolated galaxies to galaxy pairs. Therefore, there are a total of 57 individual galaxies included in sample.
 
\begin{itemize}

\item 
A total of 810 star-forming clumps are identified in our sample of 48 LIRGs, 751 of which are extra-nuclear clumps and 59 are nuclei. The number of clumps per galaxy ranges from 
two bright nuclear regions with no extra-nuclear clumps to a galaxy pair that has 95 extra-nuclear clumps. The median number of extra-nuclear clumps detected per LIRG is 15.6.

\item Star-forming clumps are resolved in our sample galaxies with effective radii of 89~pc -- 880~pc and star formation rates of extra-nuclear clumps from $2 \times 10^{-3}$ -- 4.4~M$_{\odot}$yr$^{-1}$. The median age and mass of star-forming clumps are of $9\times 10^{6}$~yr and $\sim 5 \times 10^{5}$~M$_\odot$, respectively.
The sizes and SFRs of the clumps in GOALS galaxies span the range between the regions found in local normal galaxies and those seen at higher redshifts (z = 1--3). Large and luminous star-forming clumps, similar to those seen at high-redshift, are found in the local Universe in LIRGs.

\item Star-forming clumps in LIRGs, while exhibiting overall luminosities one to two orders of magnitude higher than star-forming clumps in most local galaxies, also have a shallower clump luminosity function slope.
The GOALS sample shows a shallower clump LF slope ($-1.49\pm0.06$) than that found for star-forming regions in local, normal, non-merging galaxies ($-1.90\pm0.0$). This suggests an over abundance of luminous star-forming clumps compared to most local galaxies.
\item 
Star-forming clumps in local LIRGs span a similar range in star formation rate and size as those in high-resolution, hydrodynamic simulations (MassiveFIRE) of evolving spiral galaxies at z$\sim$2. If the clumps modeled by the MassiveFIRE simulations are similar to those found in GOALS, this might imply high gas fractions ($>$ 50\%) and extremely large star formation efficiencies in many GOALS star-forming clumps.
%Most of the GOALS clumps overlap in SFR-mass space with MassiveFIRE clumps that have intermediate to high clump gas fractions ( $>$ 50\%).  
%The simulated clumps with the highest star formation rates, lowest gas fractions, and highest star formation efficiencies appear rare in the GOALS data.
\end{itemize}

With upcoming $HST$ and ALMA observations, We will expand this sample to include galaxies with with higher molecular gas fractions (above 25\%) and study the molecular gas contents of individual clumps to place them on common star formation law relations to determine if the physical conditions in the luminous clumps are inducing exceptionally high star formation efficiencies.

%--------------------------------------------

%--------------------------------------------
\section*{Acknowledgements}
%--------------------------------------------
KLL would like to thank the anonymous referee for their useful comments as well as P. Hopkins, J. Lee, and Y. Guo for insightful discussions that all greatly improved this paper.
KLL was supported by NASA through grants HST-GO-13690.002-A and HST-GO-15241.002-A from the Space Telescope Science Institute, which is operated by the Association of Universities for Research in Astronomy, Inc., under NASA contract NAS5-26555.
T.D-S. acknowledges support from ALMA-CONICYT project 31130005 and FONDECYT regular project 1151239. G.C.P. acknowledges support from the University of Florida.
S.T.L. was supported by the NRAO Grote Reber Dissertation Fellowship. A.S.E. was supported by NSF grant AST 1109475 and by NASA through grants HST-GO1-0592.01-A, HST-GO1-1196.01-A, and HST-GO1-3364 from the Space Telescope Science Institute, which is operated by the Association of Universities for Research in Astronomy, Inc., under NASA contract NAS5-26555. V.U acknowledges funding support from the University of California Chancellor’s Postdoctoral Fellowship and NSF grant AST-1412693. Support for AMM is provided by NASA through Hubble Fellowship grant HST-HF2-51377 awarded by the Space Telescope Science Institute, which is operated by the Association of Universities for Research in Astronomy, Inc., for NASA, under contract NAS5-26555.
This research has made use of the NASA/IPAC Extragalactic Database (NED) which is operated by the Jet Propulsion Laboratory, California Institute of Technology, under contract with the National Aeronautics and Space Administration. This research also used astrodendro, a Python package to compute dendrograms of Astronomical data (http://www.dendrograms.org/).

%--------------------------------------------
%--------------------------------------------
%\vspace{5mm}

%\end{multicols}
%\newpage
%--------------------------------------------
\appendix 

%--------------------------------------------
\section{Comparison of Astrodendro and CLUMPFIND routines}

\begin{figure}[!t]
	\centering
	\includegraphics[width=0.5\textwidth]{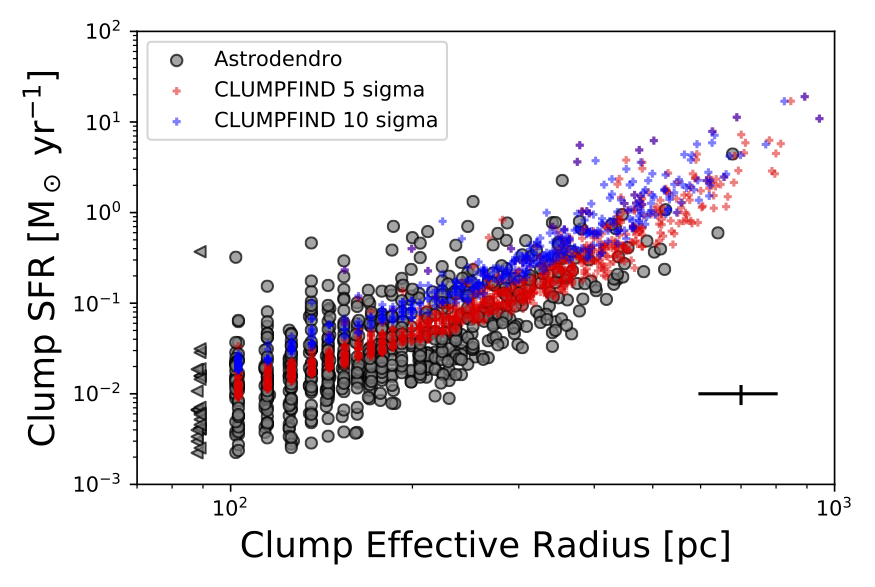}
	\caption{Direct comparison of the results of the \textsc{clumpfind} and Astrodendro (see text for details) algorthims on 10 LIRGs from GOALS. Grey circles are clumps found using Astrodendro. Plus symbols are clumps identified using \textsc{clumpfind} using a 10 sigma cutoff (in blue) and a 5 sigma cutoff (in red).  As discussed in the text, \textsc{clumpfind} fails to find the full range in clump luminosities at a given radius.  Bright clumps are shifted to larger radii, while faint clumps are boosted in luminosity.}
	\label{FIG:clumpfind}
\end{figure}

Since the high and low-z studies do not use a common algorithm to identify and measure star-forming clumps, it is important to compare how the selection might affect the measured distribution of properties.  While we have used astrodendro to identify clumps in our GOALS sources, many of the high-z studies use \textsc{clumpfind} (Williams, J. et al. 1994).  In order to test how these two algorithms might deliver different results, we have run \textsc{clumpfind} on 10 of our GOALS galaxies for a direct comparison.  When this is done, we find that while there is no systematic offset in SFR or radius for all the clumps, \textsc{clumpfind} does not effectively recover the full range of SFR at a given clump radius.
At each radius, \textsc{clumpfind} identifies a relatively narrow range of SFRs, set by the threshold for detection against the background (see Figure \ref{FIG:clumpfind}).  Since \textsc{clumpfind} does not perform a local background subtraction before measuring the radius and brightness of each clump, and above the set threshold all pixels are assigned to a clump, there is a tendency for faint clumps to be too bright, and too large. 
This effect is evident in the GOALS systems because our $HST$ near-infrared images are typically much deeper than the IFU data used to measure the high-z starbursts, and the local images contain measurable diffuse ionized gas, which must be subtracted to calculate the full range of clump properties, especially at the faint end.  

\end{document}